\documentclass[%
twocolumn,superscriptaddress
 reprint,
superscriptaddress,
 amsmath,amssymb,
 aps,
 pra,
floatfix,
]{revtex4}
\usepackage{graphicx}
\usepackage{dcolumn}
\usepackage{bm}
\usepackage{amsmath}    
\usepackage{color}      

\begin{document}


\title{Robust digital optimal control on IBM's quantum computers}

\author{Meri Harutyunyan}
\affiliation{Laboratoire Interdisciplinaire Carnot de Bourgogne, UMR CNRS 6303,  Universit\'e de Bourgogne Franche-Comt\'e, BP 47870, F-21078 Dijon, France}%
\author{Fr\'ed\'eric Holweck}
\affiliation{Laboratoire Interdisciplinaire Carnot de Bourgogne, UMR CNRS 6303,  Universit\'e de Bourgogne Franche-Comt\'e, BP 47870, F-21078 Dijon, France}%
\affiliation{Department of Mathematics and Statistics, Auburn University, Auburn, AL, USA}
\author{Dominique Sugny}
\affiliation{Laboratoire Interdisciplinaire Carnot de Bourgogne, UMR CNRS 6303,  Universit\'e de Bourgogne Franche-Comt\'e, BP 47870, F-21078 Dijon, France}%
\author{St\'{e}phane Gu\'{e}rin}
\email{sguerin@u-bourgogne.fr}
\affiliation{Laboratoire Interdisciplinaire Carnot de Bourgogne, UMR CNRS 6303,  Universit\'e de Bourgogne Franche-Comt\'e, BP 47870, F-21078 Dijon, France}%

\date{\today}

\begin{abstract}
The ability of pulse-shaping devices to generate accurately quantum optimal control is a strong limitation to the development of quantum technologies. We propose and demonstrate a systematic procedure to design robust digital control processes adapted to such experimental constraints. We show to what extent this digital pulse can be obtained from its continuous-time counterpart. A remarkable efficiency can be achieved even for a limited number of pulse parameters. We experimentally implement the protocols on IBM's quantum computers for a single qubit, obtaining an optimal robust transfer in a time $T=382$ ns.

\end{abstract}

\maketitle

\noindent \textit{Introduction.}
Operating quantum technologies requires fast and accurate control of quantum systems~\cite{Nielsen,TrainingSC}. Optimal control allows one to drive in principle the dynamics at the quantum speed limit~\cite{Caneva}. In a two-level system, one can show that it coincides with a resonant square $\pi$-pulse~\cite{Boscain}. However, the final result depends on the quality of the pulse: deviations of the amplitude (or of the time of interaction), referred to as pulse inhomogeneities, of order $\varepsilon$ leads to an error of the same order for quantum gate and of order $\varepsilon^2$ for population transfer. Improving robustness is thus vital to provide ultrahigh-fidelity quantum information processing.

Adiabatic passage technique allows such robustness but at the cost of slow and inexact dynamics~\cite{Loy,APtopology,STIRAP}. Composite pulses, made of a $\pi-$pulse sequence with well-defined phases, are a powerful and simple tool for robust control~\cite{Wimperis,Cummins,CHIRP,Jones,Vitanov2,Vitanov3} even if they are not optimal in terms of control time or energy used. Composite pulses have been recently implemented on IBM's computers~\cite{vitanov2022}. Robust optimal control protocols have been derived using gradient-based algorithms (such as GRAPE~\cite{GRAPE,kozbar2012}) or directly from the Pontryagin's Maximum Principle~\cite{Extended,boscain2021}. Alternative methods based on inverse engineering geometric control have been also recently proposed \cite{Barnes,Dridi}, where the controls are determined from a shaped trajectory in the dynamical parameter space, for instance on the Bloch sphere. The latter, referred to as robust inverse optimization (RIO)~\cite{Dridi,Laforgue2}, is derived from shortcut to adiabaticity techniques \cite{odelin2019,robustNJP,SSSP}. It combines the robustness of composite pulses, formulated as constraint integrals for a given order, with time-optimality. In the case of pulse inhomogeneities, the control has constant amplitude but features a time-continuous phase modulation.

Robustness to variations in system parameters is not the only key limitation to the experimental implementation of control protocols. Another hurdle relates to the ability of pulse-shaping devices to generate the theoretically designed  driving process. With the notable exception of composite pulses, the majority of studies has assumed that the amplitude or phase (or both) of the control can vary continuously over time. While this assumption is natural in experiments where arbitrary waveform generators are available, this procedure is problematic when the hardware is limited, for example, to specific pulse shapes and thus to a finite number of control parameters. At this stage, a numerical solution can be found by starting from the most general pulse shape and optimizing the corresponding free parameters~\cite{skinner}. However, this approach is not ideal because it generally prevents any analytical or geometric study and it makes numerical optimization more difficult by increasing the number of traps in the control landscape~\cite{rabitz2012}.

We propose in this Letter a general procedure to overcome these two limitations, demonstrated in the case of a one-qubit robust control. In view of an experimental proof of principle, we consider below specifically the case of the RIO technique on IBM's quantum computers~\cite{IBMQ}, based on superconducting transmon qubits. We stress that the same approach could be used with other optimized control and to more complex systems, but this example has the advantage of highlighting that geometric properties can still be used in this constrained control problem. Instead of producing continuous time-dependent phase, which is a relatively difficult practical task in such systems~\cite{SPQ},
we develop a simpler and efficient digital version (referred to as DRIO). Digital control has been originally proposed and explored in the context of adiabatic passage with a series of femtosecond pulses~\cite{Shapiro} and in the context of stimulated Raman passage, adiabatic \cite{Vaitkus} or not \cite{Paparelle}, with application in spatial adiabatic passage \cite{Vaitkus2}. In DRIO, the control phase is varied digitally (or in a piecewise way), produced with a short pulse series, where each pulse constituent features a constant phase. This shares similarity with composite pulses but with the important difference that one uses subpulses of small area (much smaller than $\pi$): The total pulse area is optimal (i.e. time-optimal for a given peak amplitude), and thus always smaller by construction than its composite counterpart for the same robustness order.


\textit{Theory of digital control.}
We consider a two-state system driven by a train of resonant pulses of delay $\tau$, where each individual pulse $n$ features the same shape $0\le\Lambda(t)\le1$, of characteristic duration $\sigma$, but has arbitrary phase $\varphi_n$ and peak amplitude $\Omega_n$. The two-state Hamiltonian reads in the rotating wave approximation (RWA) and in the basis of bare states
(in units such that $\hbar=1$)
\begin{equation}
\label{model}
\hat H(t)=\frac{1}{2}\sum_{n} \Omega_n \mathbf{1}_{n}(t) \Lambda\Bigl(\frac{t-t_n}{\sigma}\Bigr)
\left[\begin{array}{cc}
0 & e^{-i\varphi_n} \\
 e^{i\varphi_n} & 0\end{array}\right]
\end{equation}
with pulses centered in $t_n=n\tau$.
We denote the indicator function on the interval $I_n$:
$\mathbf{1}_{n}(t)=1\text{ if t$\in I_n\equiv
[t_{n-\frac{1}{2}},t_{n+\frac{1}{2}}[$};\  0 \text{ otherwise}$.

We consider smooth pulses, such as for instance Gaussian of the form  $\Lambda(t)=e^{-t^2}$,
that do not overlap, 
$\sigma\ll\tau$. Typically, one can take $\tau=6\sigma$ for the Gaussian pulses. In that
situation, one can safely omit the indicator function from Hamiltonian \eqref{model}
to which we associate the propagator $\hat U(t,t_i)$ (with $t_i$ the initial time).

The integer $n$ labeling the pulses in the train runs in principle from
$-\infty$ to $+\infty$, i.e. $n=0$ corresponds to the central pulse, and the peak Rabi frequency of the pulse $n$
is constructed from a continuous shape $0\le\Pi(t)\le1$ that is sampled as
\begin{equation}
\label{Rabin}
\Omega_n=\Omega_0 \Pi(t_{n}/T),
\end{equation}
where $T$ denotes its characteristic duration.
One can choose for instance a smooth Gaussian shape $\Pi(t)=e^{-t^2}$ for implementing digital adiabatic passage, or a constant (square) pulse $\Pi(t)=1$ as it will be the case for DRIO.
The phases $\varphi_n$ form a piecewise constant function whose value varies from pulse to
pulse with the same timescale as the peak Rabi frequency
\begin{equation}
\label{Phasen}
\varphi_n=\varphi(t_n)\equiv \widetilde\varphi(t_n/T).
\end{equation}
We show below that this digital model \eqref{model} can be made approximately equivalent to an effective continuous Hamiltonian, see \eqref{DefHr}. The general strategy is thus to derive the peak Rabi frequencies $\Omega_n$ and the phases $\varphi_n$ from the analysis of the continuous model.

Since the phase is constant for each pulse, one can alternatively consider the Hamiltonian
\begin{equation}
\hat H_\delta(t)=\frac{1}{2}\sum_{n}A_n
\delta\left(t-t_n\right)\left[\begin{array}{cc}
0 & e^{-i\varphi_n} \\
 e^{i\varphi_n} & 0\end{array}\right],
\end{equation}
associated to the propagator $\hat U_\delta(t,t_i)$,
where the Gaussian pulse area
$A_n=\Omega_n\int_{t_{n-\frac{1}{2}}}^{t_{n+\frac{1}{2}}}
dt\,\Lambda\left(\frac{t-t_n}{\sigma}\right) =\sqrt{\pi}\sigma\Omega_n$
between $t_{n\pm\frac{1}{2}}=(n\pm\frac{1}{2})\tau$
is concentrated into a
Dirac $\delta$ pulse. One can
indeed connect the propagators
\begin{subequations}
\begin{align}
&\hat U\bigl(t_{n+\frac{1}{2}},t_{n-\frac{1}{2}}\bigr)=
\hat U_\delta\bigl(t_{n+\frac{1}{2}},t_{n-\frac{1}{2}}\bigr)
=\hat U_\delta(t_n^+,t_n^-)\\
&=\left[\begin{array}{cc}\cos (A_n/2) & -ie^{-i\varphi_n}\sin (A_n/2)\\
-ie^{i\varphi_n}\sin (A_n/2) & \cos (A_n/2)\end{array}\right]
\end{align}
\end{subequations}
with $t_n^{\pm}$ denoting times immediately before and after $t_n=n\tau$, respectively. We highlight the fact that there is no additional phase coming from the timeshifts
$t_{n+\frac{1}{2}}\to t_n^+$ and $t_{n-\frac{1}{2}}\to t_n^-$ since the diagonal elements of the Hamiltonian are zero.


In the next step, we incorporate the phases in the wavefunction applying the piecewise constant transformation
\begin{equation}
T(t)=\left[\begin{array}{cc} e^{-i\sum_n\mathbf{1}_{n}(t)\varphi_n/2} & 0\\
0 & e^{i\sum_n\mathbf{1}_{n}(t)\varphi_n/2}\end{array}\right],
\end{equation}
which leaves unchanged the population of the states.
We obtain
\begin{align}
H_\delta&\equiv T^{\dagger}\hat H_\delta T-iT^{\dagger}\frac{dT}{dt} = \notag\\
&=\frac{1}{2}\sum_{n}
\left[\begin{array}{cc}
-\delta\bigl(t-t_{n-\frac{1}{2}}\bigr)\Delta\varphi_n
 &\delta\left(t-t_n\right)A_n \\
\delta\left(t-t_n\right)A_n
& \delta\bigl(t-t_{n-\frac{1}{2}}\bigr)\Delta\varphi_n\end{array}\right]
\label{HamT}
\end{align}
with
\begin{equation}
\Delta\varphi_n=\varphi_n-\varphi_{n-1}
\end{equation}
and the corresponding solution
$\phi(t)=T^{\dagger}(t)\Phi(t)$
with $\Phi(t)$ the state solution of the original problem (but with the Dirac
$\delta$ pulses): $\Phi(t_n^+)=\hat U_\delta\left(t_n^+,t_n^-\right)\Phi(t_n^-)$, i.e.
\begin{align}
\Phi(t)=\hat U_{\delta}(t,t_i)\Phi(t_i)=T(t) U_{\delta}(t,t_i) T^{\dagger}(t_i) \Phi(t_i).
\end{align}
The Hamiltonian~\eqref{HamT} is characterized by the superposition of two time-shifted trains of coupling and detuning terms.



When $\Omega_n\equiv\Omega_0$ and $\varphi_n\equiv\varphi_0$
are constant for any $n$, i.e. $\Delta\varphi_n=0$, the interaction is strictly periodic if one considers an infinite number of pulses and the Hamiltonian reads
\begin{equation}
\label{DefKperiodic}
H_\delta=\frac{A_0}{2\tau}\left[\begin{array}{cc}
0 & 1\\
1
& 0\end{array}\right]
\sum_{k}e^{ik\gamma t}
\end{equation}
with the frequency of the pulse repetition
$\gamma=2\pi/\tau$.
We have here used the
Poisson formula for the Dirac distribution leading to the spectral representation of the Dirac comb, 
$\sum_{n}\delta\left(t-n\tau\right)=\frac{1}{\tau}\sum_{k}
e^{ik \gamma t}$.
When $\Omega_n$ and $\varphi_n$ vary as functions of $n$,
one can use the sampling property, $\sum_{n}f(n\tau)\delta\left(t-n\tau\right)=
f(t)\sum_{n}\delta\left(t-n\tau\right),$ for a function $f(t)$,
which leads to the Hamiltonian
\begin{align}
\label{DefH}
&H_\delta
=\frac{1}{2\tau}\sum_{k}
\left[\begin{array}{cc}
-  e^{i\pi k}\Delta\varphi(t) &A(t) \\
A(t) & e^{i\pi k}\Delta\varphi (t) \end{array}\right]
e^{ik\gamma t}
\end{align}
such that
$\Delta\varphi\bigl(t_{n-\frac{1}{2}}\bigr)=\Delta\varphi_n,\ A(t_n)=A_n$.
If one considers piecewise constant peak Rabi frequencies of the form~\eqref{Rabin}, it is
natural to take for Gaussian subpulses
\begin{equation}
\label{DefA}
A(t)=\sqrt{\pi}\sigma\Omega_0 \Pi(t/T).
\end{equation}
For the detuning, one can determine a function from backward and forward Taylor expansions
of the phase~\eqref{Phasen} at time $t_{n-\frac{1}{2}}$
\begin{align}
\varphi_n&=\varphi(t_n)=\varphi(t_{n-\frac{1}{2}})+\frac{\tau}{2} \varphi'(t_{n-\frac{1}{2}})+\cdots\nonumber\\
\varphi_{n-1}&=\varphi(t_{n-1})=\varphi(t_{n-\frac{1}{2}})-\frac{\tau}{2} \varphi'(t_{n-\frac{1}{2}})+\cdots\nonumber
\end{align}
i.e.
\begin{align}
\label{Dphi}
&\Delta\varphi(t_{n-\frac{1}{2}})=\tau\varphi^{\prime}(t_{n-\frac{1}{2}})+\frac{1}{3}\left(\frac{\tau}{2}\right)^3
\varphi^{\prime\prime\prime}(t_{n-\frac{1}{2}})+\cdots\nonumber\\
&+\frac{2}{(2p+1)!}\left(\frac{\tau}{2}\right)^{2p+1}
\varphi^{(2p+1)}(t_{n-\frac{1}{2}})+\cdots
\end{align}
where $p\ge0$ is an integer.
At this stage, no approximation has been made in Hamiltonian~\eqref{DefH}, whatever the
variations of $\Delta\varphi(t)$ and $A(t)$ are, when one considers an infinite number of pulses, i.e. $T\gg\tau$.
The timescales of the digital control
need thus to satisfy 
\begin{equation}
\sigma\ll\tau\ll T.
\end{equation}
The Hamiltonian~\eqref{DefH} is not
piecewise anymore, but it contains an infinite number of modes with
the same amplitude $A(t)/2\tau$. The mode $k=0$ is near resonant while the other ones
can be considered as perturbation in the weak-field limit.
At the lowest approximation, one only takes into account 
the resonant term $k=0$, that we refer to as the second RWA. It can be applied when
\begin{equation}
\label{CondRWA2}
A(t)\ll \sqrt{\vert 4\pi^2 -\Delta\varphi(t)^2\vert }.
\end{equation}
This leads to the resonant effective Hamiltonian,
derived from~\eqref{DefH} for $k=0$
\begin{equation}
\label{DefHr}
H_{\delta,0}=
\frac{1}{2\tau}
\left[\begin{array}{cc}
- \Delta\varphi (t)&A(t) \\
A (t)& \Delta\varphi(t) \end{array}\right] \equiv
\frac{1}{2}
\left[\begin{array}{cc}
- \Delta &\Omega \\
\Omega & \Delta \end{array}\right].
\end{equation}
This effective Hamiltonian includes an effective
detuning $\Delta\equiv\Delta\varphi(t)/\tau$ and an effective Rabi frequency
corresponding to the original piecewise Rabi frequency area divided by $\tau$, 
$\Omega\equiv A(t)/\tau$.
This effective Hamiltonian~\eqref{DefHr} can be analysed from standard
techniques.
For instance, if one considers adiabatic passage, it is more efficient
when $\int A(t) dt\equiv\sqrt{\pi}\sigma\Omega_0T\int\Lambda(s)ds\gg\tau$, and for a sufficiently
slow evolution of the detuning.

\begin{figure}[tb]\centering
  \includegraphics[width=1.05\columnwidth]{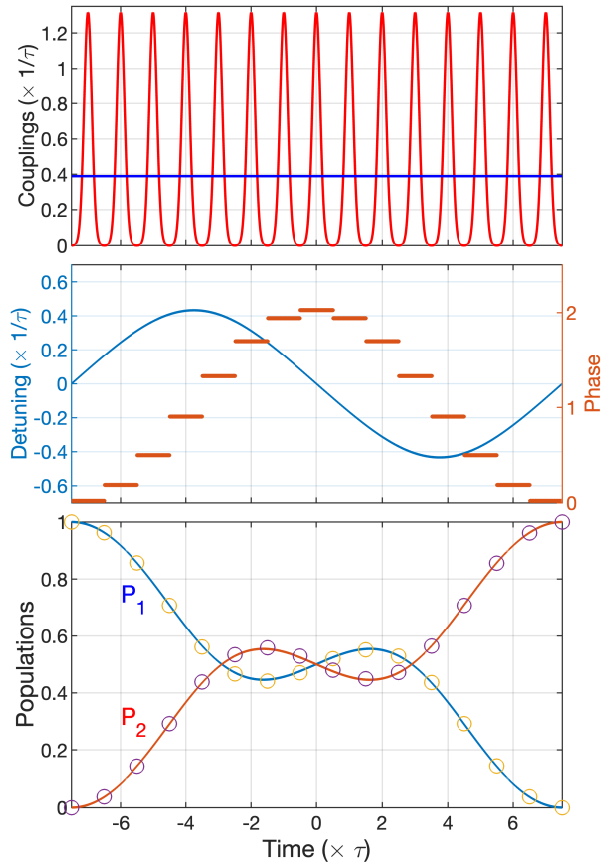}
  \caption{(Color online)
Digital and continuous third-order RIO with $N=15$. $\tau$ is the delay between the pulses and the full duration is $T=N\tau$. Upper frame: the Rabi frequency pulse series of peak amplitude $\Omega_0$ and the corresponding continuous constant Rabi frequency $\Omega$. Middle frame: the digital (piecewise) phases and the corresponding detuning. Lower frame: Population dynamics from the effective continuous model~\eqref{DefHr} (full line) and at the beginning and the end of each constituent pulse of the digital model~\eqref{model} (circles), showing complete population transfers (exactly for continuous RIO, and with an error of less than $10^{-4}$ for DRIO). The total pulse area is $1.86\pi$ for both models.
}
\label{fig1}
\end{figure}
\begin{figure}[tb]\centering
  \includegraphics[width=1.05\columnwidth]{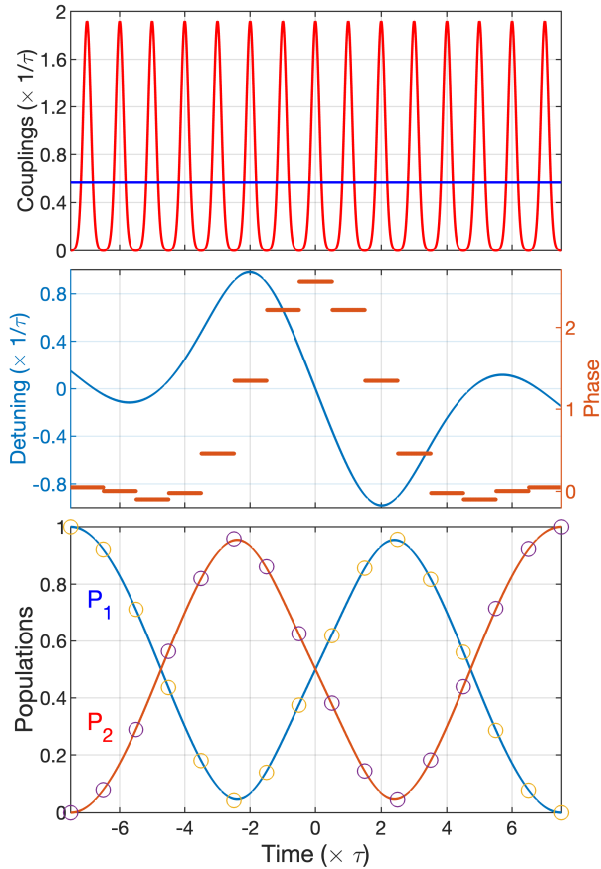}
  \caption{(Color online)
Same as Fig.~\ref{fig1}, but for the fifth-order RIO and DRIO, of total pulse area $2.71\pi$. 
}
\label{fig2}
\end{figure}

\noindent\textit{Digital robust inverse optimization}.
We consider robust inverse optimization (RIO), where robustness is imposed with respect to pulse inhomogeneities (i.e. amplitude and/or width) and optimization with respect to pulse duration, i.e in minimum time (for a given pulse peak). At  third order, the solution is a constant pulse of amplitude $\Omega$ and duration $T=1.86\pi/\Omega$ and a detuning of Jacobi elliptic cosine form
\begin{align}
\label{Dcn_}
\Delta=\Delta_0\,\text{cn}\left(\omega (t-t_i)+K(m),m\right),\ t\in[0,T]
\end{align}
with, the initial time $t_i$, and, for the problem of complete population transfer,
$m=0.235$, $\omega=1.149\Omega$, and $\Delta_0=1.114\Omega$.

The peak amplitudes of the pulses are constant $\Omega_n=\Omega_0$, since given by \eqref{Rabin} where $\Pi(t)=1$, with the Rabi coupling $\Omega_0$ given by \eqref{DefA}, $\Omega_0 =\tau\Omega/\sqrt{\pi}\sigma$. We consider a large number $N$ of pulses, typically $N=15$, giving the full duration $T=N\tau$, i.e.  $\Omega_0 =T\Omega/\sqrt{\pi}N\sigma $. We use for the delay  $\tau=6\sigma$, i.e.  $\Omega_0 =6T\Omega/\sqrt{\pi}N\tau$. The phase $\varphi(t)$ is solution of \eqref{Dphi}, extended at all times:
\begin{align}
\Delta&=\varphi^{\prime}+\frac{\tau^2}{24}
\varphi^{\prime\prime\prime}+\cdots= \varphi^{\prime}[1 +O((\tau/T)^2)] \approx \varphi^{\prime}.
\end{align}
It is next digitalized at times $t_n$ when the pulses reach their peak value.

Figure~\ref{fig1} shows the dynamics of the continuous and digital RIO for $N=15$. DRIO achieves a very good accuracy with an error in population transfer less than $10^{-4}$. A time-optimal protocol of fifth-order robustness with respect to pulse inhomogeneities can be derived by following the techniques described in Ref.~\cite{Dridi}. 
It gives a constant pulse of amplitude $\Omega$ and duration $T=2.71\pi/\Omega$. The dynamics are shown in Fig.~\ref{fig2}.

\noindent\textit{Experimental results.}
We performed the experiments of the robustness profiles using IBM Quantum Experience~\cite{IBMQ}. It is built with superconducting transmon qubits, which can be controlled by microwave pulses. We used the low-level quantum computing Qiskit Pulse \cite{QiskitPulse}, as a module of the open-source framework Qiskit \cite{Qiskit}. The processor used
is {\em ibmq\_manila}, which is one of the IBM five-qubit Falcon Processors (Falcon r5.11L). Note that one obtained the same  final result  for the other five-qubit systems of the IBM Quantum Experience.

\begin{figure}[tb]\centering
  \includegraphics[width=1.05\columnwidth]{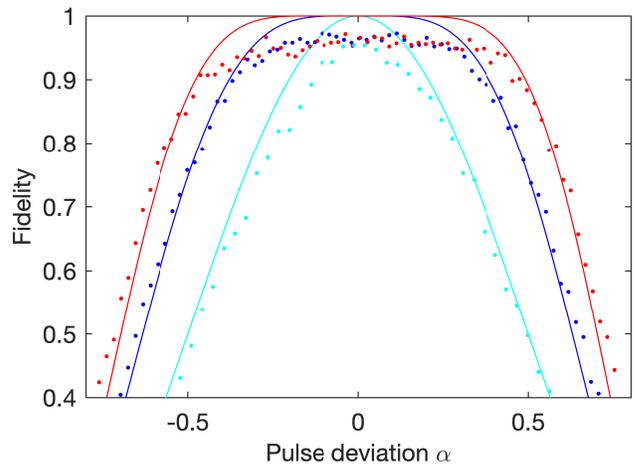}
  \caption{(Color online)
Experimental (dotted lines) digital $\pi$-pulse (light blue), third- (dark blue) and fifth-order (red) DRIO robustness profile as a function of the relative pulse amplitude deviation $\alpha$, compared to their theoretical predictions (full lines).
}
\label{fig3}
\end{figure}

The parameters of the qubit, calibrated at the time of the experiment, are the following: qubit transition frequency of 4.971~GHz, anharmonicity of -0.34378~GHz, $T_1$ and $T_2$ coherence times of 153.89~$\mu$s and 46.19~$\mu$s, respectively, and readout error of $3\%$ (all are averaged values).
We performed each experiment applying a sequence of Gaussian pulses with the appropriate phase shifts, according to Fig. \ref{fig1} and \ref{fig2}, respectively,  where each single pulse has a duration of $\sigma = 3\times\sqrt{2}$~ns. The total duration of the full process is $T=382$ ns.
In our figures we have plotted the transition probability as a function of the deviation $-1\le\alpha\le1$ from the theoretical optimal value, $\Omega\to\Omega(1 + \alpha)$, with $\Omega=1.86\pi/T$ (i.e. 15.3 MHz) and $\Omega=2.71\pi/T$ (i.e. 22.3 MHz) for the third-order and fifth-order robustness, respectively. Each point corresponds to an experiment repeated 1024 times. We have checked that the experimental result does not depend on the ratio $\sigma/\tau$  and slightly on the repetition number that can vary from 1024 to 8096 times.\\

The experimental profiles of the single (square) Rabi pulse, the third- and fifth-order DRIO are shown in Fig.~\ref{fig3}. A reasonable match between theory and experiment is achieved.
The small discrepancy amounts to approximately $3\%$, which corresponds  to the measurement error, as published by IBM at the time of the experiment.
Apart this error, the excitation profile fits remarkably well the theoretical prediction; the efficiency and robustness of DRIO are clearly demonstrated.

\noindent\textit{Conclusions.} We focus in this study on the experimental implementation of a simple digital optimization control procedure for realizing a robust one-qubit state-to-state transfer based on its time-continuous derivation. This approach has the decisive advantage of being able to account for the limitations of pulse-shaping devices, while using to some extent geometric and numerical properties of time-continuous optimal robust control protocols. We have shown the remarkable accuracy of the digital control, compatible with quantum information ultrahigh fidelity, with a digitalization featuring a number of subpulses as low as 15. The next step of this study could be the implementation of one-qubit quantum gates~\cite{garon2013} and the control of coupled qubits~\cite{schulte2005}. In particular, the use of a non-constant pulse amplitude can be similarly implemented by modulating the peaks of the subpulses according to the continuous control. This digital control protocol opens the experimental implementation of optimal driving strategies in complex systems.

\begin{acknowledgments}
We acknowledge support from the EUR-EIPHI Graduate School (17-EURE-0002) and from the European Union's Horizon 2020 research and innovation program under the Marie Sklodowska-Curie Grant No. 765075 (LIMQUET). We acknowledge the use of
IBM Quantum services for this work; the views expressed are those of the authors, and do not reflect the
official policy or position of IBM.
\end{acknowledgments}

\end{document}